\documentclass[aps,pre,twocolumn,showpacs]{revtex4}

\begin{document}

\title{Underlying mechanism of numerical instability in large eddy 
simulation of turbulent flows}

\author{Masato Ida}
\email[E-Mail: ]{ida@icebeer.iis.u-tokyo.ac.jp}
\homepage[URL:]{http://ktolab.iis.u-tokyo.ac.jp/ida/index.htm}
\altaffiliation[Present address: ]{Center for Promotion of Computational Science and Engineering, Japan Atomic Energy Research Institute, 6-9-3 Higashi-Ueno, Taito-ku, Tokyo 110-0015, Japan; E-Mail: ida@koma.jaeri.go.jp.}
\affiliation{Collaborative Research Center of Frontier Simulation Software for Industrial Science, Institute of Industrial Science, University of Tokyo, 4-6-1 Komaba, Meguro-Ku, Tokyo 153-8505, Japan
}

\author{Nobuyuki Taniguchi}
\affiliation{Information Technology Center, University of Tokyo, 2-11-16 Yayoi, Bunkyo-Ku, Tokyo 113-8658, Japan
}

\begin{abstract}
This paper extends our recent theoretical work concerning the feasibility of 
stable and accurate computation of turbulence using a large eddy simulation 
[Ida and Taniguchi, Phys.~Rev.~E {\bf 68}, 036705 (2003)]. In our 
previous paper, it was shown, based on a simple assumption regarding the 
instantaneous streamwise velocity, that the application of the Gaussian 
filter to the incompressible Navier-Stokes equations can result in the 
appearance of a numerically unstable term that can be decomposed into 
positive and negative viscosities. That result raises the question as to 
whether an accurate solution can be achieved by a numerically stable 
subgrid-scale model. In the present paper, based on assumptions regarding 
the statistically averaged velocity, we present similar theoretical 
investigations to show that in several situations, the shears appearing in 
the statistically averaged velocity field numerically destabilize the 
fluctuation components because of the derivation of a numerically unstable 
term that represents negative diffusion in a fixed direction. This finding 
can explain the problematic numerical instability that has been encountered 
in large eddy simulations of wall-bounded flows. The present result suggests 
that this numerical problem is universal in large eddy simulations, and that 
if there is no failure in modeling, the resulting subgrid-scale model can 
still have unstable characteristics; that is, the known instability problems 
of several existing subgrid-scale models are not something that one may 
remove simply by an artificial technique, but must be taken seriously so as 
to treat them accurately.
\end{abstract}

\pacs{47.11.+j, 47.27.Eq, 47.10.+g, 83.85.Pt}

\maketitle

\section{Introduction}
\label{sec1}
Turbulence is one of the unsolved problems of physics \cite{refa1}. 
Because a 
complete theoretical description has not yet been achieved even for a 
relatively simple flow configuration, numerical simulations are commonly 
used as a powerful tool for analyzing turbulent flows. Typical numerical 
approaches include the direct numerical simulation (DNS), the 
Reynolds-averaged Navier-Stokes (RANS), and the large eddy simulation (LES). 
In DNS, all the scales of motion in the turbulent flows are resolved by 
sufficiently fine computational grids, whereas in RANS, only the evolution 
of mean quantities is solved. LES is an intermediate technique between these 
approaches, directly solving the large scales but modeling small-scale 
eddies by employing a subgrid-scale (SGS) model (or a subfilter-scale model) 
that approximately accounts for the effects of the small scales on the large 
scales \cite{ref1}. Because LES enables us to solve time-dependent large-scale 
turbulent flow problems with a relatively small computational time and 
storage compared to those required for DNS, this technique has recently been 
used not only for academic studies but also for practical industrial flow 
computations that need time-dependent solutions.

One of the major problems of LES is numerical instability. As is already 
known, several existing SGS models (e.g., the tensor-diffusivity \cite{ref2}, 
the scale-similarity \cite{ref3}, and the dynamic Smagorinsky \cite{ref4} models), which have 
been constructed based on a filtering procedure and the statistical 
properties of turbulence, have a numerically unstable property, and hence 
some artificial numerical treatments (e.g., smoothing or clipping of the SGS 
stress) have been incorporated so as to guarantee numerical stability \cite{ref1,ref4,ref5,ref6,ref7}. While the mechanisms of these models' unstable behaviors have been described in the literature (e.g., Refs.~\cite{ref1,ref8,ref9,ref24,ref7}), the underlying 
reason as to why the SGS models have an unstable property has, to the 
authors' knowledge, not yet been fully clarified. To construct an excellent 
SGS model that is free from artificial, unphysical numerical treatments and 
has applicability to a wide range of flow configurations with high accuracy 
and robustness, it is necessary to elucidate the underlying mechanism of 
those unstable properties. Is this numerical problem caused by a failure in 
modeling or by other factors? To answer this question, it should be 
meaningful to consider a similar but idealized question {\it whether a completely accurate SGS model, if it exists, would be numerically stable}.

In Ref.~\cite{ref24}, Leonard has shown that the tensor-diffusivity model, 
which was 
derived by truncating an exact expansion series of the SGS stress terms and 
is thus exact under a certain condition, should behave unstably along the 
stretching directions of fluid motion. This unstable behavior results from 
the negative diffusivity of that model, which makes the governing equations 
ill-conditioned and leads to numerical instability when treated numerically 
by, e.g., finite difference methods. Winckelmans et al.~\cite{ref7} have performed 
several numerical experiments using the pure (and mixed) tensor-diffusivity 
model and have pointed out that the model behaves unstably under a 
wall-bounded flow condition, while it can provide a stable result for 
turbulent isotropic decay. In a recent paper \cite{ref10} we have shown 
theoretically that the filtering procedure, which is the most fundamental 
component of LES, is a potential seed of the numerical instability in LES. 
Without having resorted to any SGS model, but based on a simple 
assumption regarding the streamwise component of flow velocity, we found 
that under a wall-bounded flow condition, the application of a Gaussian 
filter (one of the filters commonly used in LES \cite{ref11}) to the Navier-Stokes 
equations results in the appearance of a numerically unstable term, a 
cross-derivative of the filtered velocity component, which can be decomposed 
into diffusion and negative-diffusion terms. The above findings indicate 
that apparently the application of Gaussian filtering stabilizes the 
physical properties of the governing equations, but this is not always the 
case. That result implies that even a completely accurate SGS model that 
perfectly reproduces the physical properties of the true SGS components 
might not always be numerically stable.

However, both Leonard's and Ida and Taniguchi's theoretical works are 
insufficient to fully clarify the mechanism of numerical instabilities in 
actual simulations. While those studies have only considered the 
instantaneous nature of the SGS model or of the SGS forces, it is possible 
that their time-averaged nature is dissipative and that stable simulations 
can thus be achieved. Indeed, as mentioned above, the tensor-diffusivity 
model, which has an unstable property, can provide stable solutions for an 
isotropic turbulent flow \cite{ref7}. This observation suggests that further efforts 
must be made to improve our understanding of the numerical instabilities in 
LES.

In the present paper, we have extended our previous work so as to obtain 
more acceptable and reliable results and to show that this numerical problem 
is universal in LES based on Gaussian filtering. We have previously assumed 
that the instantaneous value of the streamwise velocity component is 
linearly proportional to the distance from a plane wall parallel to the bulk 
flow. In contrast, in the present paper our discussion assumes that the 
statistically averaged streamwise velocity component is linearly 
proportional to the distance from the wall, an assumption that is more 
realistic because this is the case of a viscous sublayer forming near a 
plane wall. Furthermore, we discuss cases where an axisymmetric swirl exists 
in the statistically averaged velocity field. As will be shown in the 
following section, a numerically unstable term with a time-independent 
coefficient can appear in both situations. This term is always unstable in a 
fixed range of directions, while the previously discussed unstable 
characteristics of the model and of the SGS terms depend on the directions 
of instantaneous stretching \cite{ref24} or of instantaneous shears \cite{ref10}. Therefore, 
the present theoretical result can more accurately explain the numerical 
instabilities frequently encountered in inhomogeneous flows that involve a 
strong steady shear in the mean velocity field.

This paper is organized as follows: In Sec.~\ref{sec2:2} we reexamine the 
wall-bounded flow case, and in Sec.~\ref{sec2:3} we extend our theory to swirling 
flow cases. In the appendixes, we provide additional notes on further 
extensions of our theory. Section \ref{sec3} presents concluding remarks.

\section{Theoretical investigations}

\subsection{Filtering approach}
Incompressible, viscous fluid flows are described by the Navier-Stokes 
equations, which read:
\begin{equation}
\label{eq1}
\frac{\partial u_i }{\partial t}+\frac{\partial u_j u_i }{\partial x_j 
}=-\frac{\partial p}{\partial x_i }+\nu \frac{\partial ^2u_i }{\partial x_j 
\partial x_j }\quad \mbox{for}\;i=1,2,3,
\end{equation}
\begin{equation}
\label{eq2}
\frac{\partial u_i }{\partial x_i }=0,
\end{equation}
where Einstein's summation convention is assumed, and $u_i =u_i (x_1 ,x_2 
,x_3 ,t)$ are the velocity components, $p$ is the pressure divided by the 
constant fluid density, and $\nu $ is the kinematic viscosity. In LES, a 
filter is applied to this system of equations to separate the large and 
small scales. This filtering procedure is in general achieved by the 
following convolution:
\begin{equation}
\label{eq3}
\bar {F}(x,\ldots ,t)=\int\limits_{X=-\infty }^{X=\infty } {L(x-X)F(X,\ldots 
,t)dX} ,
\end{equation}
where $\overline {(\cdot )} $ denotes the filtered quantity, $x$ is an 
independent variable of an arbitrary function $F$, and $L(X)$ is the filter 
function. In the present study, we assume $L(X)$ to be the Gaussian function
\begin{equation}
\label{eq4}
L(X)=\sqrt {\frac{\gamma }{\Delta ^2\pi }} \exp \left( {-\frac{\gamma 
X^2}{\Delta ^2}} \right),
\end{equation}
which satisfies $\int_{X=-\infty }^{X=\infty } {L(X)} dX=1$, where $\Delta $ 
is the filter width (assumed to be constant), and $\gamma $ is a real, 
positive constant. In the previous paper we set to $\gamma =1/2$ as done by 
Klimas for the Vlasov equation, while $\gamma =6$ is generally used in LES 
\cite{ref11}; we adopt the latter value in the present study. Appling the filtering 
operation to Eqs.~(\ref{eq1}) and (\ref{eq2}) yields
\begin{equation}
\label{eq5}
\frac{\partial \bar {u}_i }{\partial t}+\frac{\partial \overline {u_j u_i } 
}{\partial x_j }=-\frac{\partial \bar {p}}{\partial x_i }+\nu \frac{\partial 
^2\bar {u}_i }{\partial x_j \partial x_j }
\end{equation}
or
\begin{equation}
\label{eq6}
\frac{\partial \bar {u}_i }{\partial t}+\frac{\partial \bar {u}_j \bar {u}_i 
}{\partial x_j }=-\frac{\partial \bar {p}}{\partial x_i }+\nu \frac{\partial 
^2\bar {u}_i }{\partial x_j \partial x_j }-\frac{\partial \tau _{ij} 
}{\partial x_j },
\end{equation}
\[
\tau _{ij} \equiv \overline {u_i u_j } -\bar {u}_i \bar {u}_j ,
\]
with
\begin{equation}
\label{eq7}
\frac{\partial \bar {u}_i }{\partial x_i }=0,
\end{equation}
where we used
\[
\overline {\left( {\frac{\partial f}{\partial t}} \right)} =\frac{\partial 
\bar {f}}{\partial t}\quad \mbox{and}\quad \overline {\left( {\frac{\partial 
f}{\partial x_j }} \right)} =\frac{\partial \bar {f}}{\partial x_j }
\]
($f$ being a dependent variable), and $\tau _{ij} $ is the so-called SGS 
stress tensor that generally needs to be modeled.

Before getting into the main subject, we present here some mathematical 
tools useful for the present investigation. Supposing that the Gaussian 
filter is applied in the $x_i $ direction, we have
\begin{equation}
\label{eq8}
\overline {(x_i f)} =x_i \bar {f}+\frac{\Delta ^2}{2\gamma }\frac{\partial 
\bar {f}}{\partial x_i }
\end{equation}
and
\begin{equation}
\label{eq9}
\overline {(x_j f)} =x_j \bar {f}\quad \mbox{for}\;j\ne i,
\end{equation}
which can be given in the same manner as shown in Refs.~\cite{ref12,ref13,ref10}, and 
have been utilized in, e.g., Ref.~\cite{ref14}. Using Eq.~(\ref{eq8}), we have
\begin{equation}
\label{eq10}
\bar {x}_i =x_i ,
\end{equation}
\begin{equation}
\label{eqad1}
\overline {(x_i ^2)} =x_i ^2+\frac{\Delta ^2}{2\gamma }.
\end{equation}

Based on these results, in the following subsections we perform theoretical 
investigations concerning the Gaussian filtered Navier-Stokes equations 
under a plane channel and swirling flow conditions.

\subsection{Plane channel flow}
\label{sec2:2}
Suppose that $x_1 $, $x_2 $, and $x_3 $, respectively, are the streamwise, 
the wall-normal, and the spanwise directions, and a plane solid wall is set 
for $x_2 \le 0$. Decomposing the velocity components into the statistically 
averaged value $U_i $ and fluctuation part $u'_i $, we have
\begin{equation}
\label{eq11}
u_i ({\rm {\bf x}},t)=U_i ({\rm {\bf x}})+u'_i ({\rm {\bf x}},t),
\end{equation}
where $U_i $ is assumed to not depend on time. Furthermore, we assume that
\begin{eqnarray}
\label{eq12}
{\rm {\bf U}}({\rm {\bf x}}) &=& (U_1 ({\rm {\bf x}}),U_2 ({\rm {\bf x}}),U_3 
({\rm {\bf x}})) \nonumber \\ 
 &=& (\beta x_2 ,0,0)
\end{eqnarray}
with $\beta $ being a real constant, i.e., the mean streamwise velocity is 
linearly proportional to the distance from the wall. (The extension to more 
general cases is briefly discussed in Appendix \ref{secA}.) Based on these 
assumptions, one knows
\begin{equation}
\label{eq13}
\frac{\partial U_i }{\partial x_i }=\frac{\partial u'_i }{\partial x_i 
}=0.
\end{equation}

Substituting Eqs.~(\ref{eq11}) and (\ref{eq12}) into the convection terms in the equation 
for $u_1 $ and using Eq.~(\ref{eq13}), we have
\begin{eqnarray}
\label{eq14}
\frac{\partial u_j u_1 }{\partial x_j } &=& U_1 \frac{\partial u'_1 
}{\partial x_1 }+\frac{\partial U_1 }{\partial x_2 }u'_2 +\frac{\partial 
u'_j u'_1 }{\partial x_j } \nonumber \\ 
 &=& \beta x_2 \frac{\partial u'_1 }{\partial x_1 }+\beta u'_2 
+\frac{\partial u'_j u'_1 }{\partial x_j }.
\end{eqnarray}
Appling Gaussian filtering in the wall-normal direction to this equation 
yields
\begin{eqnarray}
\label{eq15}
\frac{\partial \overline {u_j u_1 } }{\partial x_j } &=& \bar {U}_1 
\frac{\partial \bar u'_1 }{\partial x_1 }+\beta \frac{\Delta ^2}{2\gamma 
}\frac{\partial ^2\bar u'_1 }{\partial x_1 \partial x_2 }+\frac{\partial 
\bar {U}_1 }{\partial x_2 }\bar u'_2 +\frac{\partial \overline {u'_j 
u'_1 } }{\partial x_j } \nonumber \\ 
 &=& \beta \bar {x}_2 \frac{\partial \bar u'_1 }{\partial x_1 }+\beta 
\frac{\Delta ^2}{2\gamma }\frac{\partial ^2\bar u'_1 }{\partial x_1 
\partial x_2 }+\beta \bar u'_2 +\frac{\partial \overline {u'_j u'_1 
} }{\partial x_j }, \nonumber \\ 
{}
\end{eqnarray}
where we use Eq.~(\ref{eq8}) and $U_1 =\bar {U}_1 $ given easily by Eq.~(\ref{eq10}). Even 
if a three-dimensional Gaussian filter is adopted, almost the same formula 
is derived because $U_1 $ only depends on $x_2 $, and hence no additional 
term is derived by filtering in the other directions. Meanwhile, the 
convection terms in terms of the filtered velocity components, $\partial 
(\bar {u}_j \bar {u}_1 )/\partial x_j $, can be written as
\begin{equation}
\label{eq16}
\frac{\partial \bar {u}_j \bar {u}_1 }{\partial x_j }=\bar {U}_1 
\frac{\partial \bar u'_1 }{\partial x_1 }+\frac{\partial \bar {U}_1 
}{\partial x_2 }\bar u'_2 +\frac{\partial \bar u'_j \bar u'_1 
}{\partial x_j }.
\end{equation}
From Eqs.~(\ref{eq15}) and (\ref{eq16}), we have
\begin{equation}
\label{eq17}
\frac{\partial \tau _{1j} }{\partial x_j }=\beta \frac{\Delta ^2}{2\gamma 
}\frac{\partial ^2\bar u'_1 }{\partial x_1 \partial x_2 }+\frac{\partial 
(\overline {u'_j u'_1 } -\bar u'_j \bar u'_1 )}{\partial x_j }.
\end{equation}
Substituting this and $\partial ^2\bar {U}_1 /\partial x_j \partial x_j =0$ 
into Eq.~(\ref{eq6}) yields
\begin{eqnarray}
\label{eq18}
\frac{\partial \bar {u}_1 }{\partial t}+\frac{\partial \bar {u}_j \bar {u}_1 
}{\partial x_j }&=&-\frac{\partial \bar {p}}{\partial x_1}+\nu \frac{\partial 
^2\bar u'_1 }{\partial x_j \partial x_j }-\beta \frac{\Delta ^2}{2\gamma 
}\frac{\partial ^2\bar u'_1 }{\partial x_1 \partial x_2 } \nonumber \\
&&-\frac{\partial 
(\overline {u'_j u'_1 } -\bar u'_j \bar u'_1 )}{\partial x_j },
\end{eqnarray}
which can be considered an equation for both $\bar {u}_1 $ and $\bar 
u'_1 $ because
\[
\frac{\partial \bar {u}_1 }{\partial t}=\frac{\partial \bar u'_1 
}{\partial t}.
\]
The second-to-the-last term of Eq.~(\ref{eq18}) is a cross-derivative of the 
dependent variable $\bar u'_1 $. As has been proved in several studies 
(e.g., Refs.~\cite{ref13,ref10}), the derivatives of this type should be numerically 
unstable when, for instance, a finite difference technique is adopted for 
solving the equation. In what follows, we consider the stability condition 
for Eq.~(\ref{eq18}).

The last three terms of Eq.~(\ref{eq18}) may make a dominant contribution to the 
numerical stability of LES. We will first focus our attention on the second- 
and third-to-the-last terms:
\begin{equation}
\label{eq19}
\nu \frac{\partial ^2\bar u'_1 }{\partial x_j \partial x_j }-\beta 
\frac{\Delta ^2}{2\gamma }\frac{\partial ^2\bar u'_1 }{\partial x_1 
\partial x_2 }.
\end{equation}
Appling the coordinate transformation of
\begin{equation}
\label{eq20}
(\xi ,\eta )=\left( {\frac{x_1 +x_2 }{\sqrt 2 },\frac{-x_1 +x_2 }{\sqrt 2 }} 
\right),
\end{equation}
which represents rotation by $45^\circ $ around the $x_3 $ axis, Eq.~(\ref{eq19}) is 
rewritten as
\begin{equation}
\label{eq21}
\nu \left( {\frac{\partial ^2{\bar {u}}'_1 }{\partial \xi ^2}+\frac{\partial 
^2{\bar {u}}'_1 }{\partial \eta ^2}+\frac{\partial ^2{\bar {u}}'_1 
}{\partial x_3 ^2}} \right)-\beta \frac{\Delta ^2}{2\gamma }\left( 
{\frac{1}{2}\frac{\partial ^2{\bar {u}}'_1 }{\partial \xi 
^2}-\frac{1}{2}\frac{\partial ^2{\bar {u}}'_1 }{\partial \eta ^2}} \right).
\end{equation}
From this, it can be seen that the following restriction is required to 
ensure that Eq.~(\ref{eq21}) is a positive viscosity:
\begin{equation}
\label{eq22}
\nu -\frac{\Delta ^2}{4\gamma }\left| \beta \right|\ge 0.
\end{equation}

The shear gradient $\beta $ may have different values depending on the 
problems considered. We consider here a simple example to show concretely 
how restriction (\ref{eq22}) works in a realistic situation. As is well known, in 
the viscous sublayer of turbulent channel flows, the statistically averaged 
streamwise velocity follows \cite{ref15}
\begin{equation}
\label{eq23}
U_1 (x_2 )=u_\tau \left( {\frac{\left| {u_\tau } \right|}{\nu }x_2 } 
\right)
\end{equation}
with $u_\tau $ being the wall-friction velocity. Using Eq.~(\ref{eq23}) and 
rewriting reduce Eq.~(\ref{eq22}) to
\begin{equation}
\label{eq24}
\frac{\left| {u_\tau } \right|}{\nu }\Delta =\Delta ^+\le 2\sqrt \gamma ,
\end{equation}
where the superscript + denotes a quantity in the wall units. For $\gamma =6$, 
this result essentially corresponds to the stability condition derived by 
Kobayashi and Shimomura from the tensor-diffusivity term in a dynamic SGS 
model \cite{ref9}.

We should note here that Eq.~(\ref{eq24}) is a minimum condition for stabilizing the 
numerical solution; the last term of Eq.~(\ref{eq18}), which we have neglected, can 
also be a seed of numerical instability because if, for example, $u'_1 
\propto x_2 $ is true at a certain instant and location, then
\[
\overline {\left( {u'_1 \frac{\partial u'_1 }{\partial x_1 }} \right)} 
-\bar u'_1 \frac{\partial \bar u'_1 }{\partial x_1 }\propto 
\frac{\partial ^2\bar u'_1 }{\partial x_1 \partial x_2 }.
\]
(See Ref.~\cite{ref10}.) Nevertheless, this incomplete condition creates a strong 
restriction on the grid width ($h)$. The grid width should satisfy $h^+\le 
\sqrt 6 \approx 2.4$ for $\Delta ^+=2h^+$, a setting that has frequently 
been used, or $h^+\le \sqrt 6 /2\approx 1.2$ for $\Delta ^+=4h^+$, which is 
necessary for the contribution of the SGS force to be significant compared 
to the truncation and aliasing errors of a second-order finite difference 
scheme \cite{ref16}. (We should note here that this restriction is imposed only 
on the grid width in the wall-normal direction because only the filtering in 
this direction yields the cross-derivative term, as mentioned already.) This 
result implies that at least several grid points are needed in the viscous 
sublayer to guarantee numerical stability, which is a formidable restriction 
in practical applications of a large-scale high-Reynolds-number turbulent 
flow. It is known that when the inner layer of turbulent boundary-layer 
flows is resolved, the number of grid points required for LES exceeds 
current computational capacities even at moderate Reynolds numbers, and 
hence developments of SGS models that are applicable to a coarse wall-normal 
grid and of wall models that do not need a strong refinement of the 
near-wall grids are the most pressing issues of LES; see a recent review by 
Piomelli and Balaras \cite{ref27}. The numerical problem that we have suggested 
above provides a taste of the difficulties of constructing an excellent 
model.

The stability condition presented above would be partly weakened if the last 
term of Eq.~(\ref{eq18}) is modeled using an eddy-viscosity model. However, this 
treatment does not always achieve a stable solution. Actually, Winckelmans 
et al.~\cite{ref7} have observed by numerical experiments that for a plane channel 
flow, a dynamic mixed model based on the tensor-diffusivity model (which can 
accurately represent the characteristics of second-order cross derivatives 
\cite{ref9}) generates numerical instabilities that eventually make the numerical 
simulation blow up. To remove those instabilities, they were forced to add 
an artificial damping coefficient to the tensor-diffusivity portion. In the 
mixed model that they used, the coefficient of eddy-viscosity part was well 
tuned by a dynamic procedure. Essentially the same observations of the 
numerical instability of the dynamic mixed model can be found in Ref.~\cite{ref9} by 
Kobayashi and Shimomura. In those studies, the wall-normal grid width was 
set to be several times larger than that used in DNS. That is, those studies 
suggest that, for a large grid (and filter) width, which is much-needed in 
practical applications, the addition of an eddy-viscosity term does not 
always guarantee numerical stability. If, on the other hand, all the SGS 
terms are modeled using an eddy-viscosity model whose coefficient is 
guaranteed to be positive, the numerical stability of the filtered system 
would be guaranteed in many cases. However, it has been suggested by many 
researchers that eddy-viscosity models have several deficiencies that make 
the LES results suspicious (e.g., Refs.~\cite{ref1,ref32,refa3,ref33,ref8,ref20,ref34,ref37}). For 
instance, as has frequently been demonstrated by {\it a priori} tests using DNS 
techniques, the SGS quantities modeled using eddy-viscosity models do not 
correlate well with the actual SGS quantities \cite{ref33,ref1,ref8,ref20,ref25}, implying 
that these models have low accuracy and/or narrow applicability. Indeed, in 
the present case the eddy-viscosity models cannot describe the 
cross-derivative term in Eq.~(\ref{eq18}), which is a hybrid of positive and 
negative diffusions. Moreover, eddy-viscosity models were constructed under 
the assumptions that the subgrid turbulence is isotropic and the filter 
scale is in an inertial range, assumptions that are in general both violated 
in inhomogeneous turbulent flows, especially near walls. Furthermore, the 
dynamic types of eddy-viscosity models need artificial techniques for 
smoothing the model coefficient and for clipping its negative values \cite{ref1,ref4,ref5,ref6,ref29}. Based on the above, we consider the eddy-viscosity models to be 
insufficient. Most of the recent efforts at SGS modeling have been devoted 
to developing a model that does not strongly rely on eddy-viscosity models 
\cite{ref28,ref30,ref25,ref31,ref7,ref6,ref22,ref37}. Models based on a kind of ``defiltering'' procedure \cite{ref28,ref30,ref37} are one such approach, determining significant 
portions of the SGS terms by an analytical, not empirical, procedure. The 
mixed models based on the tensor-diffusivity term \cite{ref31,ref7,ref6} can also be 
categorized into this type \cite{ref25}. We should note here that attempts to 
improve eddy-viscosity models are also being continued; see, e.g., Refs.~\cite{ref29,ref34,ref35,ref21}.

It is worth noting that the numerical stability of the cross-derivative term 
in Eq.~(\ref{eq18}) is time-independent, as this term always reveals a negative 
diffusivity in a fixed range of directions, whereas the numerical stability 
of the last term of the same equation may be time-dependent. As mentioned in 
Sec.~\ref{sec1}, Winckelmans et al.~have pointed out that the numerical 
stability of 
the tensor-diffusivity model depends on the problems that they have 
considered; obtaining stable results is difficult to achieve for a strongly 
inhomogeneous (wall-bounded) flow, even when an eddy-viscosity term is 
added, but is possible for a homogeneous flow even by the pure 
tensor-diffusivity model \cite{ref7}. They have explained these results as follows: 
For the homogeneous flow, the direction in which the tensor-diffusivity term 
reveals the negative diffusivity evolves continuously in time and space, and 
hence the model term may be dissipative on time average. However, in the 
inhomogeneous flow case, long-lived negative diffusion events can take place 
near the wall, leading to numerical instability and a resulting divergence 
of the solution. As mentioned above, those numerical findings indicate an 
inadequacy of the previous stability analyses by Leonard \cite{ref24} and by Ida and 
Taniguchi \cite{ref10}, which have only considered the instantaneous behavior of the 
SGS model or the SGS term. The present theoretical findings, showing the 
appearance of a term that is always numerically unstable in fixed directions 
and hence is unstable on time-average as well, can more correctly explain 
the problematic numerical instability encountered in the simulations of 
plane channel flows \cite{ref7,ref9}.

\subsection{Swirling flow}
\label{sec2:3}
Turbulent flows involving a large-scale swirl are of practical importance in 
connection with combustion engineering, aeroacoustics, meteorology, and so 
on. In this subsection, we discuss the LES of turbulent flows having a 
rotating mean velocity about the $x_3 $ (or $z)$ axis, and show that the 
existence of the swirl can cause numerical instability through the Gaussian 
filtering operation.

First, we consider a very simple case in which the swirl has a constant 
angular velocity; that is, the mean velocity takes the form of
\begin{equation}
\label{eq25}
{\rm {\bf U}}({\rm {\bf x}})=(\alpha x_2 ,-\alpha x_1 ,0),
\end{equation}
where $\alpha $ is a real constant. This mean velocity satisfies
\[
{\rm {\bf U}}\cdot {\rm {\bf x}}^\ast =0,
\]
\[
\left| {\rm {\bf U}} \right|/\left| {{\rm {\bf x}}^\ast } \right|=\left| 
\alpha \right|,
\]
and
\[
\frac{\partial U_i }{\partial x_i }=0,
\]
where ${\rm {\bf x}}^\ast \equiv (x_1 ,x_2 ,0)$. As a result, we know
\begin{equation}
\label{eq26}
\frac{\partial u'_i }{\partial x_i }=0.
\end{equation}
Using the above, the convection terms for $u_1 $ are rewritten as
\begin{equation}
\label{eq27}
\frac{\partial u_j u_1 }{\partial x_j }=U_1 \frac{\partial u'_1 }{\partial 
x_1 }+U_2 \frac{\partial u'_1 }{\partial x_2 }+\frac{\partial U_1 
}{\partial x_2 }u'_2 +U_2 \frac{\partial U_1 }{\partial x_2 
}+\frac{\partial u'_j u'_1 }{\partial x_j },
\end{equation}
where the first four terms on the right-hand side (RHS) result from the 
existence of the swirl, among which the first and second terms,
\begin{equation}
\label{eq28}
U_1 \frac{\partial u'_1 }{\partial x_1 }+U_2 \frac{\partial u'_1 
}{\partial x_2 }=\alpha x_2 \frac{\partial u'_1 }{\partial x_1 }-\alpha 
x_1 \frac{\partial u'_1 }{\partial x_2 },
\end{equation}
may have a significant influence on the numerical stability of the equation 
for $\bar u'_1 $.

We show here that the filtered formula and its stability change depending on 
how the Gaussian filtering is applied. Appling the Gaussian filter in the 
$x_2 $ direction to Eq.~(\ref{eq28}) results in
\[
\overline {\left( {U_1 \frac{\partial u'_1 }{\partial x_1 }} \right)} 
+\overline {\left( {U_2 \frac{\partial u'_1 }{\partial x_2 }} \right)} 
=\alpha x_2 \frac{\partial \bar u'_1 }{\partial x_1 }+\alpha 
\frac{\Delta ^2}{2\gamma }\frac{\partial ^2\bar u'_1 }{\partial x_1 
\partial x_2 }-\alpha x_1 \frac{\partial \bar u'_1 }{\partial x_2 },
\]
which can be rewritten as
\begin{equation}
\label{eq29}
\overline {\left( {U_1 \frac{\partial u'_1 }{\partial x_1 }} \right)} 
-\bar {U}_1 \frac{\partial \bar u'_1 }{\partial x_1 }+\overline {\left( 
{U_2 \frac{\partial u'_1 }{\partial x_2 }} \right)} -\bar {U}_2 
\frac{\partial \bar u'_1 }{\partial x_2 }=\alpha \frac{\Delta 
^2}{2\gamma }\frac{\partial ^2\bar u'_1 }{\partial x_1 \partial x_2 }.
\end{equation}
The cross-derivative term appearing on the RHS may cause numerical 
instability when the absolute value of its coefficient is sufficiently 
large. In contrast, when the Gaussian filter is applied in both the $x_1 $ 
and $x_2 $ directions, we obtain a stable formula,
\begin{eqnarray*}
\overline {\left( {U_1 \frac{\partial u'_1 }{\partial x_1 }} \right)} 
+\overline {\left( {U_2 \frac{\partial u'_1 }{\partial x_2 }} \right)} 
&=& \alpha x_2 \frac{\partial \bar u'_1 }{\partial x_1 }+\alpha 
\frac{\Delta ^2}{2\gamma }\frac{\partial ^2\bar u'_1 }{\partial x_1 
\partial x_2 } \\
&&-\alpha x_1 \frac{\partial \bar u'_1 }{\partial x_2 
}-\alpha \frac{\Delta ^2}{2\gamma }\frac{\partial ^2\bar u'_1 }{\partial 
x_1 \partial x_2 } \\
&=& \alpha x_2 \frac{\partial \bar u'_1 }{\partial x_1 }-\alpha x_1 
\frac{\partial \bar u'_1 }{\partial x_2 },
\end{eqnarray*}
and thus,
\begin{equation}
\label{eq30}
\overline {\left( {U_1 \frac{\partial u'_1 }{\partial x_1 }} \right)} 
-\bar {U}_1 \frac{\partial \bar u'_1 }{\partial x_1 }+\overline {\left( 
{U_2 \frac{\partial u'_1 }{\partial x_2 }} \right)} -\bar {U}_2 
\frac{\partial \bar u'_1 }{\partial x_2 }=0.
\end{equation}
Equation (\ref{eq30}) has no cross derivative.

Next, we consider the same example but in the cylindrical coordinate 
$(r,\theta ,z)$. In this case, the mean velocity (\ref{eq25}) and the instantaneous 
velocity, respectively, are represented by
\begin{eqnarray}
\label{eq31}
{\rm {\bf U}}(r,\theta ,z) = (U_r ,U_\theta ,U_z ) = (0,-\alpha r,0),
\end{eqnarray}
and
\begin{equation}
\label{eq32}
(u_r ,u_\theta ,u_z )=(u'_r ,U_\theta +u'_\theta ,u'_z ),
\end{equation}
and the nonlinear terms in the Navier-Stokes equation for $u_\theta $ take 
the form of
\begin{equation}
\label{eq33}
u_r \frac{\partial u_\theta }{\partial r}+\frac{u_\theta }{r}\frac{\partial 
u_\theta }{\partial \theta }+u_z \frac{\partial u_\theta }{\partial 
z}+\frac{u_r u_\theta }{r}\equiv h_\theta ,
\end{equation}
where the velocity components should satisfy
\begin{eqnarray}
\label{eq34}
0 &=& \frac{\partial u_r }{\partial r}+\frac{u_r }{r}+\frac{1}{r}\frac{\partial 
u_\theta }{\partial \theta }+\frac{\partial u_z }{\partial z} \nonumber \\ 
 &=& \frac{\partial u'_r }{\partial r}+\frac{u'_r 
}{r}+\frac{1}{r}\frac{\partial u'_\theta }{\partial \theta }+\frac{\partial u'_z }{\partial z}.
\end{eqnarray}
Substituting Eq.~(\ref{eq32}) into Eq.~(\ref{eq33}) yields
\begin{eqnarray}
\label{eq35}
h_\theta &=& \frac{U_\theta }{r}\frac{\partial u'_\theta }{\partial \theta 
}+\left( {\frac{\partial U_\theta }{\partial r}+\frac{U_\theta }{r}} 
\right)u'_r \nonumber \\
&& +\left( {u'_r \frac{\partial u'_\theta }{\partial 
r}+\frac{u'_\theta }{r}\frac{\partial u'_\theta }{\partial \theta 
}+u'_z \frac{\partial u'_\theta }{\partial z}+\frac{u'_r u'_\theta 
}{r}} \right).
\end{eqnarray}
The terms in the last parentheses represent the nonlinear interaction 
between the fluctuation components, whereas the remaining terms describe the 
interaction between the mean and fluctuation portions. Appling the Gaussian 
filter in the $r$ direction with a filter width of $\Delta _r $ to the first 
term results in
\begin{equation}
\label{eq36}
\overline {\left( {\frac{U_\theta }{r}\frac{\partial u'_\theta }{\partial 
\theta }} \right)} =-\alpha \frac{\partial \bar u'_\theta }{\partial 
\theta }=\frac{\bar {U}_\theta }{r}\frac{\partial \bar u'_\theta 
}{\partial \theta },
\end{equation}
where we used Eq.~(\ref{eq10}) and assumed $r\gg \Delta _r $ so that the value of 
the Gaussian filter is sufficiently close to zero at $r=0$. No unstable term 
appears in Eq.~(\ref{eq36}). Even when the Gaussian filter is applied in both the 
$r$ and $\theta $ directions, almost the same formula is obtained because 
$U_\theta $ depends only on $r$. Specifically, in the present case the 
resulting formula is not, unlike the previous case, dependent on how the 
Gaussian filter is applied.

The above results, which reveal that the numerical stability of LES depends 
not only on the flow configuration but also on the filtering strategy and 
coordinate system adopted, can be interpreted as follows: As has been shown 
in Ref.~\cite{ref10} and in the preceding subsection, the existence of a shear leads 
to the appearance of a numerically unstable term. Although no shear exists 
in the mean flow profile described by Eq.~(\ref{eq25}), a virtual shear is observed 
when the Gaussian filter is applied only in one direction in the Cartesian 
coordinate ($x_1 $ or $x_2 )$, resulting in the derivation of the 
cross-derivative term. This problem does not arise for the cylindrical 
coordinate, since the mean convection velocity in the $(r,\theta )$ space,
\[
\left( {\frac{dr}{dt},\frac{d\theta }{dt}} \right)=\left( {U_r 
,\frac{U_\theta }{r}} \right)=(0,-\alpha ),
\]
is uniform.

Lastly, we consider a case where the mean velocity has a second-order term, 
i.e., assuming
\begin{equation}
\label{eq37}
{\rm {\bf U}}(r,\theta ,z)=(0,\alpha _1 r+\alpha _2 r^2,0),
\end{equation}
where $\alpha _1 $ and $\alpha _2 $ are real constants. Substituting this 
into the first term on the RHS of Eq.~(\ref{eq35}) and applying the Gaussian filter 
in the $r$ direction, we have
\begin{equation}
\label{eq38}
\overline {\left( {\frac{U_\theta }{r}\frac{\partial u'_\theta }{\partial 
\theta }} \right)} =\left( {\alpha _1 +\alpha _2 r} \right)\frac{\partial 
\bar u'_\theta }{\partial \theta }+\alpha _2 \frac{\Delta _r ^2}{2\gamma 
}\frac{\partial ^2\bar u'_\theta }{\partial r\partial \theta },
\end{equation}
which can be rewritten as
\begin{equation}
\label{eq39}
\overline {\left( {\frac{U_\theta }{r}\frac{\partial u'_\theta }{\partial 
\theta }} \right)} -\frac{\bar {U}_\theta }{r}\frac{\partial \bar 
u'_\theta }{\partial \theta }=-\alpha _2 \frac{\Delta _r ^2}{2\gamma 
r}\frac{\partial \bar u'_\theta }{\partial \theta }+\alpha _2 
\frac{\Delta _r ^2}{2\gamma }\frac{\partial ^2\bar u'_\theta }{\partial 
r\partial \theta }
\end{equation}
because, based on Eqs.~(\ref{eq10}) and (\ref{eqad1}),
\[
\bar {U}_\theta =\alpha _1 r+\alpha _2 \left( {r^2+\frac{\Delta _r 
^2}{2\gamma }} \right).
\]
The last term of Eq.~(\ref{eq39}) is a cross derivative of the dependent variable 
and should destabilize the numerical solution, whereas the 
second-to-the-last term represents convection in the $\theta $ direction 
that can be solved stably. For $\alpha _2 =0$, the RHS terms of Eq.~(\ref{eq39}) 
vanish, and this equation corresponds to Eq.~(\ref{eq36}). More specifically, the 
instability in this case is due to the second-order component in the mean 
velocity, which violates the uniformity of the angular velocity and thus 
represents a shear.

As in the plane-channel case, the cross-derivative terms derived above have 
a time-independent coefficient and thus should lead to numerical instability 
if an unsuitably large filter is adopted. The present results imply that 
such an unstable term can appear in many situations. In a future paper, we 
will consider more general and complicated flow configurations to create a 
generalized theory of numerical instability due to filtering operations 
\cite{ref36}.

\section{Summary and conclusion}
\label{sec3}
In summary, we have theoretically investigated the numerical instability of 
LES caused by the filtering procedure based on simple assumptions regarding 
the statistically averaged velocity, and have shown that Gaussian filtering 
yields a numerically unstable term, which always reveals a negative 
diffusivity in a fixed range of directions in cases of both plane channel 
flow and swirling flow. This conclusion confirms and extends that of our 
previous work \cite{ref10}. We anticipate that it would be interesting to examine 
the relationship between this numerical instability and the shear-induced 
(physical) instabilities in turbulence \cite{ref18,ref19}.

Furthermore, we have presented several additional findings. In 
Sec.~\ref{sec2:2}, we 
showed that in channel-flow case, this numerical problem strongly restricts 
the wall-normal grid width necessary for achieving stable and accurate 
solutions. In the academic computations of a plane channel flow using an LES 
technique, the grid width in the wall-normal direction has sometimes been 
set to almost the same as that required in DNS, and no filter is adopted in 
this direction (e.g., Refs.~\cite{ref20,ref4,ref21,ref22,refa3}); the present result 
provides a grounding for this custom. Furthermore, this result indicates a 
significant difficulty in developing an excellent SGS model that can provide 
satisfactory (i.e., not only stable but also accurate) solutions, even with 
large wall-normal filter widths, which is seriously desired in practical 
applications of high- (and even moderate-) Reynolds-number turbulence. In 
Sec.~\ref{sec2:3}, based on investigations of swirling flow cases, we pointed 
out 
that the numerical stability of the filtered equations depends not only on 
the flow configuration, but also on the dimension of the Gaussian filtering 
and the adopted coordinate system. This finding would be of particular 
importance in practical applications that use a general curvilinear 
coordinate system or unstructured grids.

In the present study (and also in our previous paper), it was assumed 
implicitly that all the spectral modes contained in the velocity field are 
fully resolved because the Gaussian filter does not cut off any modes, but 
only damps the high-frequency modes. (See Appendix \ref{secB}, which presents a 
related remark concerning cases where the spectral cutoff or the top-hat 
filter is employed, both of which eliminate certain modes.) The present 
results therefore indicate that even in such an ideal case, the {\it complete} SGS model 
(or, more precisely, the {\it complete} subfilter-scale model whose characteristic length 
is determined independently from the grid width), which can completely 
reproduce the properties of real SGS stresses, can be numerically unstable 
when treated numerically, since the complete model perfectly describes the 
characteristics of the unstable cross-derivative term. This problem poses a 
dilemma for practitioners of LES who are looking for an accurate and stable 
SGS model. To improve model accuracy, this numerical instability problem 
should be confronted. However, if a rough, artificial treatment is adopted 
for stability, the accuracy of the LES results will not be guaranteed. A 
potential approach to overcoming this difficulty is, as suggested in our 
previous paper \cite{ref10}, to construct a stable and accurate numerical solver for 
the numerically unstable term, though this would admittedly be quite a 
difficult task. (Leonard \cite{ref24} and Moeleker and Leonard \cite{ref26} have proposed 
Lagrangian methods based on the tensor-diffusivity model and an anisotropic 
particle scheme to solve the negative-diffusion problems, and have achieved 
stable solutions for a two-dimensional scalar transport equation with known 
velocity fields. However, several issues remain to be overcome to extend 
those methods to the case of Navier-Stokes turbulence \cite{ref26}.) Further careful 
and vigorous considerations are necessary for this instability problem to be 
solved.

\begin{acknowledgments}
One of the authors (M.I.) thanks T. Tominaga, M. Miyazawa, and K. Matsuura for 
their valuable comments. This work was supported by the Ministry of 
Education, Culture, Sports, Science, and Technology of Japan 
(Monbu-Kagaku-Sho) under an IT research program ``Frontier Simulation 
Software for Industrial Science.''
\end{acknowledgments}

\appendix

\section{CLOSURE FOR HIGH-ORDER MEAN VELOCITY}
\label{secA}
As a first step toward constructing a general theory for the filtering 
instability, let us consider the closure of
\[
\overline {\left( {U_1 \frac{\partial u'_1 }{\partial x_1 }} \right)} 
\]
in the case where $U_1 $ is a general function of $x_2 $, under the 
assumption that the Gaussian filter is applied in the $x_2 $ direction. 
Using the Taylor expansion, the arbitrary function $U_1 (x_2 )$ is 
represented by a polynomial of infinite order:
\[
U_1 (x_2 )=\sum\limits_{n=0}^\infty {a_n x_2 ^n} ,
\]
where $a_n $ are real constants. Therefore, what we have to do is to 
consider the closure of
\begin{equation}
\label{eqA1}
\overline {\left( {x_2 ^n\frac{\partial u'_1 }{\partial x_1 }} \right)}
\end{equation}
with $n$ being an arbitrary positive integer. This aim is achieved as 
follows: Using Eq.~(\ref{eq8}) successively, Eq.~(\ref{eqA1}) is rewritten into a closed 
form in terms of $\bar u'_1 $, as
\begin{eqnarray}
\overline {\left( {x_2 \left( {x_2 ^{n-1}\frac{\partial u'_1 }{\partial 
x_1 }} \right)} \right)} &=& \chi \overline {\left( {x_2 ^{n-1}\frac{\partial 
u'_1 }{\partial x_1 }} \right)} \nonumber \\ 
 &=& \chi \chi \overline {\left( {x_2 ^{n-2}\frac{\partial u'_1 }{\partial 
x_1 }} \right)} \nonumber \\ 
 \cdots \nonumber \\ 
 &=& \chi ^n\frac{\partial \bar u'_1 }{\partial x_1 },
\end{eqnarray}
where
\[
\chi \equiv x_2 +\frac{\Delta ^2}{2\gamma }\frac{\partial }{\partial x_2 }.
\]
The resulting formula has derivatives of high orders, and hence stability 
analysis would be somewhat intricate.

\section{SPECTRAL CUTOFF AND TOP-HAT FILTERS}
\label{secB}
In Ref.~\cite{ref12}, Klimas showed that the Gaussian filtered Vlasov equation can 
be rewritten into a closed form in terms of the filtered distribution 
function without any approximation. Also, in Ref.~\cite{ref23} (see also Refs.~\cite{ref24,ref25}) Yeo showed that the Gaussian filtered Navier-Stokes equations are 
written in a closed form having an infinite series. Their results have been 
utilized in our study, as described in the present and previous papers. This 
Appendix is devoted to showing that when the spectral cutoff or the top-hat 
filter is adopted, it is in general impossible to analytically derive an 
exact, closed formula.

Let us consider two arbitrary velocity fields
\[
{\rm {\bf u}}^{(m)}({\rm {\bf x}})=(u_1 ^{(m)}({\rm {\bf x}}),u_2 
^{(m)}({\rm {\bf x}}),u_3 ^{(m)}({\rm {\bf x}}))\quad \mbox{for }m=1,2,
\]
whose spectra are
\[
{\rm {\bf \hat {u}}}^{(m)}({\rm {\bf k}})=(\hat {u}_1 ^{(m)}({\rm {\bf 
k}}),\hat {u}_2 ^{(m)}({\rm {\bf k}}),\hat {u}_3 ^{(m)}({\rm {\bf k}}))\quad 
\mbox{for }m=1,2,
\]
where ${\rm {\bf x}}=(x_1 ,x_2 ,x_3 )$ and ${\rm {\bf k}}$ is the wavenumber 
vector. If these velocities, at a certain instant $t_1 $, satisfy
\[
{\rm {\bf \hat {u}}}^{(1)}({\rm {\bf k}})\ne {\rm {\bf \hat {u}}}^{(2)}({\rm 
{\bf k}})\quad \mbox{for}\;\left| {\rm {\bf k}} \right|>k_c 
\]
but
\[
{\rm {\bf \hat {u}}}^{(1)}({\rm {\bf k}})={\rm {\bf \hat {u}}}^{(2)}({\rm 
{\bf k}})\quad \mbox{for}\;\left| {\rm {\bf k}} \right|\le k_c ,
\]
and
\[
\overline {{\rm {\bf u}}^{(1)}} ({\rm {\bf x}})=\overline {{\rm {\bf 
u}}^{(2)}} ({\rm {\bf x}})\quad \mbox{for all}\;{\rm {\bf x}},
\]
where $\overline {(\cdot )} $ denotes the spectral cutoff with an identical 
cutoff wavenumber $k_c $, then the {\it closed} equation will provide the same result 
after this instant [i.e., $\overline {{\rm {\bf u}}^{(1)}} ({\rm {\bf x}},t)=\overline {{\rm {\bf u}}^{(2)}} ({\rm {\bf x}},t)$ for $t>t_1 $], 
although the unfiltered velocities have different high-wavenumber modes. This 
unacceptable conclusion has resulted from the assumption that a closed 
formula exists.

If, on the other hand,
\[
{\rm {\bf u}}^{(1)}({\rm {\bf x}})={\rm {\bf u}}^{(2)}({\rm {\bf x}})+\cos 
(kx_2 ){\rm {\bf i}}
\]
is true at an instant (where $k$ is a constant wavenumber and ${\rm {\bf 
i}}$ denotes the unit vector in the $x_1 $ direction) and the wavelength of 
the cosine function in this equation is, for example, equal to the filter 
width of the top-hat filter applied (at least) in the $x_2 $ direction, then
\[
\overline {{\rm {\bf u}}^{(1)}} ({\rm {\bf x}})=\overline {{\rm {\bf 
u}}^{(2)}} ({\rm {\bf x}})\quad \mbox{for}\;\mbox{all}\;{\rm {\bf x}}.
\]
(Here, $\overline {(\cdot )} $ denotes the top-hat filtering.) As in the 
previous case, this result denies the existence of a closed formula.

The above results imply that in order for an exact closed formula to exist, 
the spectrum of the applied filter needs to have a nonzero value for 
$\left| {\rm {\bf k}} \right|\ne \infty $. The convolution product in the 
physical space using a filter function $L$,
\[
\overline {{\rm {\bf u}}^{(m)}} ({\rm {\bf x}})=\int\limits_{-\infty 
}^\infty {L({\rm {\bf x}}-{\rm {\bf \xi }}){\rm {\bf u}}^{(m)}({\rm {\bf \xi 
}})d{\rm {\bf \xi }}} ,
\]
is represented in the Fourier space by the simple multiplication of the 
spectra of $L$ and ${\rm {\bf u}}^{(m)}$:
\[
\overline {{\rm {\bf \hat {u}}}^{(m)}} ({\rm {\bf k}})=\hat {L}({\rm {\bf 
k}}){\rm {\bf \hat {u}}}^{(m)}({\rm {\bf k}}).
\]
Based on the above, we know that if ${\rm {\bf \hat {u}}}^{(1)}({\rm {\bf 
k}})\ne {\rm {\bf \hat {u}}}^{(2)}({\rm {\bf k}})$ for certain ${\rm {\bf 
k}}$ and $\hat {L}({\rm {\bf k}})$ has a nonzero value for all ${\rm {\bf 
k}}$, then $\overline {{\rm {\bf \hat {u}}}^{(1)}} ({\rm {\bf k}})\ne 
\overline {{\rm {\bf \hat {u}}}^{(2)}} ({\rm {\bf k}})$ at least for the 
certain ${\rm {\bf k}}$ and consequently $\overline {{\rm {\bf u}}^{(1)}} 
({\rm {\bf x}})\ne \overline {{\rm {\bf u}}^{(2)}} ({\rm {\bf x}})$ at least 
in some regions; more specifically, if the adopted filter function is not 
equal to zero for all ${\rm {\bf k}}$, filtering different velocity fields 
results in different filtered velocity fields.

The present conclusion is in opposition to the theoretical result of Carati 
et al.~\cite{ref25} which proved mathematically the existence of a closed formula in 
the top-hat case. Let us consider the cause of this discrepancy. In the 
study of Carati et al., it was first assumed that the generalized expansion 
series for one-dimensional filtering in the $x$ direction takes the form of 
\begin{equation}
\label{eq40}
\overline {(ab)} =\sum\limits_{r,s=0}^\infty {c_{rs} \partial _x^r \bar 
{a}\partial _x^s \bar {b}} ,
\end{equation}
where $a$ and $b$ are arbitrary, continuous, and differentiable functions of 
$x$ and $c_{rs} $ are real constants. We show here that this assumption is 
not always valid for top-hat filters. If $a$ is a sinusoidal wave whose 
wavelength is equal to the characteristic width of the applied top-hat 
filter, then $\bar {a}(x)=0$ for any value of $x$ and consequently the 
right-hand side of Eq.~(\ref{eq40}) is also equal to zero for any $x$ if 
$c_{rs}$ are finite values. However, the left-hand side of this equation is 
not necessarily zero under the present condition; if, for example, $b=a$, then 
$ab\ge 0$ for any $x$ and thus $\overline {(ab)} >0$. This contradiction is 
caused by the fact that the value of the top-hat filters can be zero in the 
Fourier space. The generating function, Eq.~(3.7) of Ref.~\cite{ref25} used to 
derive the generalized expansion series, is definable only for the filters 
that always have a nonzero value in the Fourier space as Gaussian filters, 
because that function diverges at the wavenumbers for which the filter value 
is zero. Therefore, the expansion series for top-hat filters is definable 
only if the filter width is smaller than the smallest resolved scale. When 
this restriction is not fulfilled, the expansion series should not converge.

\end{document}